\documentclass[%
pra,twocolumn, superscriptaddress,
showpacs,
amsmath,
amssymb,
]{revtex4-1}

\usepackage{graphicx}
\usepackage{dcolumn}
\usepackage{bm}

\widetext
\usepackage[utf8]{inputenc} 
\usepackage{mathrsfs}
\usepackage{graphicx}
\usepackage{amsmath,amsfonts}
\usepackage{indentfirst}
\usepackage{natbib}
\usepackage{color}

\begin{document}

\title{Quantum walks and gravitational waves}

\author{Pablo Arnault}
\email{pablo.arnault@upmc.fr}
\affiliation{LERMA, UMR 8112, UPMC and Observatoire de Paris, 61 Avenue de l'Observatoire, 75014 Paris, France}
\affiliation{Research Center of Integrative Molecular Systems (CIMoS), Institute for Molecular Science, National Institutes of Natural Sciences, 38 Nishigo-Naka, Myodaiji, Okazaki 444-8585, Japan.}

\author{Fabrice Debbasch}
\email{fabrice.debbasch@gmail.com}
\affiliation{LERMA, UMR 8112, UPMC and Observatoire de Paris, 61 Avenue de l'Observatoire, 75014 Paris, France}

\date{\today}

\begin{abstract}

A new family of discrete-time quantum walks (DTQWs) propagating on a regular (1+2)D spacetime lattice is introduced. The continuum limit of these DTQWs is shown to coincide with the dynamics of a Dirac fermion coupled to an arbitrary relativistic gravitational field. This family is used to model the influence of arbitrary linear gravitational waves (GWs) on DTQWs. Pure shear GWs are studied in detail. We show that on large spatial scales, the spatial deformation generated by the wave induces a rescaling of the eigen-energies by a certain anisotropic factor which can be computed exactly. The effect of pure shear GWs on fermion interference patterns is also investigated, both on large scales and on scales comparable to the lattice spacing.
\end{abstract}


\maketitle

\section{Introduction}

Discrete-time quantum walks (DTQWs) are unitary quantum automata. They are not stochastic but can be viewed nevertheless as formal quantum analogues of classical random walks. They were first introduced by Feynman\cite{FeynHibbs65a, Schweber86a} in the 1940's and later re-introduced, as 'quantum random walks' by Aharonov et al. \cite{ADZ93a}, and in a systematic way by Meyer \cite{Meyer96a}.
 DTQWs have been realized experimentally  with a wide range of physical objects and setups \cite{Sanders03a, Perets08a, Schmitz09a, Karski09a, Zahring10a, Schreiber10a, Sansoni11a}, and are studied in a large variety of contexts, ranging from fundamental quantum physics \cite{var96a, Perets08a} to quantum algorithmics \cite{Amb07a, MNRS07a}, solid-state physics \cite{Bose03a, Aslangul05a, Burg06a, Bose07a} and biophysics \cite{Engel07a, Collini10a}.

It has been shown recently that the continuum limit of several DTQWs defined on regular (1+1)D spacetime lattices coincides with the dynamics of Dirac fermions coupled, not only to electric fields 
\cite{AD14, AD15}, 
but also to arbitrary non-Abelian Yang-Mills gauge fields \cite{ADMDB16}
and to relativistic gravitational fields \cite{DMD13b,DMD14,Arrighi_curved_1D_15,SFGP15a}.
Gravitational waves (GWs) \cite{W84a} are of great interest, both theoretically and experimentally, and their effects on DTQWs are thus worth investigating. The interest about GWs has been renewed by their recent direct detection \cite{Abb16a}. Linear GWs ressemble electromagnetic waves. In particular, 
GWs can be expanded as a superposition of plane waves and each plane wave as a superposition of two polarization states, both polarizations being perpendicular to the direction of propagation. The effect of these plane GWs on matter is thus typically studied in the polarization plane and it makes little sense to envisage the action of GWs on (1+1)D DTQWs. Performing a valid study of how GWs influence DTQWs thus requires building DTQWs coupled to (1+2)D gravitational fields.

We start by presenting a new family of (1+2)D DTQWs whose continuum limit coincides with the dynamics of a Dirac fermion coupled to arbitrary (1+2)D gravitational fields. The construction of this family is inspired by the 
(1+1)D  construction presented in \cite{DMD14}. The DTQWs in the (1+2)D spacetime depend on two parameters, which code for the mass of the walker and for the finite spacing of the lattice, and on four time- and space-dependent angles. We then show how to choose these four angles to describe linear GWs. 
A generic linear GW on the lattice can be considered as the superposition of three waves: two compression waves along the directions of the lattice and a shear wave coupling directly two directions of the lattice through non-diagonal metric components. Shear effects are of particular interest in relativistic gravitation and are present in generic solutions of Einstein equations \cite{E11a}. We thus focus on pure shear GWs and examine in detail their action on the DTQWs. Our main results are the following. On large spatial scales, pure shear GWs rescale locally the eigen-energies by an anisotropic factor, to make up for the space deformation induced by the wave, and the eigen-polarizations are modified as well. On smaller scales 
comparable to a few lattice spacings, both polarizations and energies are modified in a non-trivial way; this has the effect of changing significantly the interference pattern of two 
fermion eigen-modes. A final section discusses the construction of the DTQWs and mentions several avenues open to further study.

%

\section{DTQWs in (1+2) dimensions} \label{sec:first}

Consider a quantum walker moving on a two-dimensional lattice (discrete space) with nodes labelled by $(p_1,p_2) \in \mathbb{Z}^2$. Let $j \in \mathbb{N}$ label discrete time and $(b_-,b_+)$ be a certain time- and position-independent basis of the two-dimensional coin Hilbert space of the walker, that we identify to $((1,0)^{\top},(0,1)^{\top})$, where  $\top$ denotes the transposition. 
The state of the walker at time $j$ and point $(p_1, p_2)$ is described by a two-component wave function $\Psi_{j,p_1,p_2} = \psi^-_{j,p_1,p_2} b_- + \psi^+_{j,p_1,p_2} b_+$. The collection $( \Psi_{j,p_1,p_2})_{(p_1,p_2) \in \mathbb{Z}^2}$ will be denoted by $\Psi_j$.

The time evolution of $\Psi_j$ is fixed by a time-dependent unitary operator $V_j$:
\begin{equation} \label{eq:walk}
\Psi_{j+1} = V_j \Psi_{j} \, .
\end{equation}
This unitary operator involves two real positive parameters, $\epsilon$ and $m$, and four time- and space-dependent angles $\theta^{11}$, $\theta^{12}$, $\theta^{21}$ and $\theta^{22}$; it consists essentially in a combination of rotations in spin space 
and of translations in physical space, along the two directions of the lattice. 
The operator $V_j$ consists in a rather complicated combination of rotations in spin space 
and of translations in physical space along the two directions of the lattice. 
It is defined as follows:
\begin{align}
V_j & = \boldsymbol{\Pi}^{-1} \Big[ \mathbf{W}_1(\theta^{12}_{j}) \mathbf{W}_2(\theta^{22}_{j}) \Big]  \boldsymbol{\Pi} \\
& \ \ \ \ \ \times
\Big[ \mathbf{W}_2(\theta^{21}_{j})\mathbf{W}_1(\theta^{11}_{j})\Big] 
\mathbf{Q} \left( \epsilon(m-T_{\epsilon} ( \boldsymbol{\theta})  /4 )\right)  , \nonumber 
\end{align}
where $\theta^{kl}_j = ( \theta^{kl}_{j,p_1,p_2})_{ (p_1,p_2) \in \mathbb{Z}^2 }$ for $(k, l ) \in \left\{ 1, 2 \right\}^2$, 
and $\boldsymbol{\theta} = ( \theta^{kl}_j)_{ (k, l, j) \in \left\{1, 2 \right\}^2 \times \mathbb{N} }$.
The operator $\boldsymbol{\Pi}$ is
\begin{equation}
\boldsymbol{\Pi}  = \frac{1}{\sqrt{2}}\left[
\begin{array}{cc}
-i & 1 \\
-1 & i
\end{array}
\right].
\end{equation}
The operators $\mathbf{W}_k(\theta_{j})$, $k \in \left\{ 1, 2 \right\}$, are defined by
\begin{align} \label{eq:1D_block}
\mathbf{W}_{k}(\theta_j) = 
\mathbf{R}^{-1}(\theta_j) \ \Big[ \mathbf{U}(\theta_j) \mathbf{S}_{k} \mathbf{U}(\theta_j) \mathbf{S}_{k} \Big] \ \mathbf{R}(\theta_j) \, ,
\end{align}
where $\mathbf{S}_{k}$ is the spin-dependent translation operator in the $k$ spatial direction, while $\mathbf{R}(\theta_j)$ and $\mathbf{U}(\theta_j)$ are rotations in the coin Hilbert space:
\begin{align}
&\left(\mathbf{S}_{1} \Psi_j \right)_{p_1, p_2} = \left[
\begin{array}{c}
\psi^{-}_{j,p_1+1,p_2} \\
\psi^{+}_{j,p_1-1,p_2}
\end{array}
\right] \nonumber  \\
&\left(\mathbf{S}_{2} \Psi_j \right)_{p_1, p_2} = \left[
\begin{array}{c}
\psi^{-}_{j,p_1,p_2 + 1} \\
\psi^{+}_{j,p_1,p_2 - 1}
\end{array}
\right] ,
\end{align}
\begin{align}
&\left( \mathbf{R}(\theta_j) \Psi_j  \right)_{p_1, p_2} = \mathbf{r} (\theta_{j, p_1, p_2}) \Psi_{j, p_1, p_2} \nonumber  \\
&\left( \mathbf{U}(\theta_j) \Psi_j  \right)_{p_1, p_2} = \mathbf{u} (\theta_{j, p_1, p_2}) \Psi_{j, p_1, p_2} \, , 
\end{align}
with 
\begin{align}
\mathbf{u}(\theta)  &= \left[
\begin{array}{cc}
 -\cos{\theta} & i \sin{\theta} \\
 -i \sin{\theta} & \cos{\theta}
\end{array}
\right] \nonumber \\
\mathbf{r}(\theta)  &=  \left[
 \begin{array}{cc}
 i \cos(\theta/2) & i \sin(\theta/2) \\
 - \sin(\theta/2) & \cos(\theta/2)
\end{array}
\right].
\end{align}
Finally
\begin{equation}
\mathbf{Q}(\mathrm{M})  = \left[
\begin{array}{cc}
\cos (2\mathrm{M}) & -i \sin (2 \mathrm{M}) \\
-i \sin (2\mathrm{M}) & \cos (2 \mathrm{M})
\end{array}
\right],
\end{equation}
and
\begin{equation} \label{eq:function_T}
T_{\epsilon} (\boldsymbol{\theta}) = \sum_{k=1}^2 \left( C^{k 2} \, D_0^\epsilon (C^{-1})^{1k} - C^{k1} \, D_0^\epsilon (C^{-1})^{2k}  \right)  ,
\end{equation}
with $(D_0^\epsilon K)_{j,p_1,p_2} = (K_{j+1,p_1,p_2} - K_{j,p_1,p_2})/\epsilon$ for any quantity $K_{j,p_1,p_2}$ defined on the spacetime lattice, and
\begin{equation}
\left[C^{kl} \right]= \left[\cos \theta^{kl} \right].
\label{eq:defE}
\end{equation}

Let us now comment on the construction of the operator $V_j$.
The new (1+2)D DTQWs defined by (\ref{eq:walk}) are
inspired by the (1+1)D DTQWs
introduced in \cite{DMD13b}. As shown in this earlier work, the coupling of a DTQW with a gravitational field in (1+1)D can only be obtained if the walker performs at each time step not one, but two jumps in the spatial direction. Thus, one would expect that coupling DTQWs with gravitational fields in (1+2)D could be obtained by letting the walker perform two jumps in each spatial direction at each time-step.  These jumps are represented by the operators $\mathbf{W}_1(\theta^{11}_{j})$ and $\mathbf{W}_2(\theta^{22}_{j})$ which we will discuss in more detail below. These $2 \times 2$ jumps however do not suffice and do not take into account the shear generated by generic gravitational fields. This shear couples directions and manifest itself by non-diagonal metric coefficients which do not vanish identically. To take the shear into account, the quantum walker has to perform two extra jumps in each direction at each time step. These two extra jumps are represented by the operators $\mathbf{W}_1(\theta^{12}_{j})$ and $\mathbf{W}_2(\theta^{21}_{j})$. 
We are interested in DTQWs whose continuum limit coincides with the Dirac equation. Choosing a given representation of the (1+2)D Clifford algebra to write down Dirac equation in explicit form introduces an apparent symmetry breaking between the two space directions $x$ and $y$, even in flat spacetime.
The operator $\boldsymbol{\Pi}$ enters the definition of the DTQWs to account for this apparent symmetry breaking.
The operators $\mathbf{W}_k(\theta_j)$ themselves deserve some more comments. Each operator involves two jumps, represented by the operators
$\mathbf{S}_{k}$'s, and two mixing operators, represented by the $\mathbf{U}(\theta_j)$'s. As shown in \cite{DMD13b} for the (1+1)D, involving only these operations would deliver the correct Dirac dynamics in the continuum limit, but not in a standard `fixed' spin basis. The operators $\mathbf{R}^{-1}(\theta_j)$ and $\mathbf{R}(\theta_j)$ compensate for this basis difference. 

The operator $\mathbf{Q} \left( \epsilon( m-T_{\epsilon} \left( \boldsymbol{\theta}\right)/4)\right)$ is included (i) to endow the DTQW with the mass $m$ (see the continuum limit presented in the next section), (ii) to provide the `mass-like' extra term $-T_{\epsilon} \left( \boldsymbol{\theta} \right)$ which is present in the (1+2)D Dirac equation (see next section and Appendix \ref{app:mass_like_Dirac}). Note that $T_{\epsilon} (\boldsymbol{\theta} )$ (i) is non local in the time $j$ and (ii) vanishes if $[C^{kl}]$ is either diagonal, antidiagonal, or independent of $j$.

It is convenient to introduce two three-dimensional objects,
$\left[e_{(a)}^\mu\right]$ and $\left[e_\mu^{(a)}\right]$, $(a, \mu) \in \{0, 1, 2\}^3$, whose spatial parts coincide respectively with $C^{kl}$ and $(C^{-1})^{kl}$. More precisely, 
we define $e^0_{(0)} = e^{(0)}_0 = 1$, $e^0_{(1)} = e^{(0)}_1 = e^1_{(0)} = e^{(1)}_0 = e^0_{(2)} = e^{(0)}_2 = e^2_{(0)} = e^{(2)}_0 = 0$ and $e^k_{(l)} = C^{kl}$, $e^{(k)}_l = (C^{-1})^{kl}$, for $(k, l) \in \{1, 2\}^2$.
These two objects allow to rewrite $T_{\epsilon}  (\boldsymbol{\theta} )$ in the more compact form (see Appendix \ref{app:mass_like} for a derivation):
\begin{equation}\label{eq:Tcomp}
T_{\epsilon}  (\boldsymbol{\theta} ) = - \varepsilon^{a b c} \eta_{cd} e_{(a)}^\mu D_b^\epsilon e^{(d)}_\mu \, ,
\end{equation}
where $[\eta_{ab}] = \mbox{diag}(1, -1, -1)$, $\varepsilon^{a b c}$ is the totally antisymmetric symbol, with $\varepsilon^{012}=1$, and the finite difference operator $(D_b^\epsilon) = (D_0^\epsilon, D_1^\epsilon, D_2^\epsilon)$ can be given arbitrary spatial components $D_1^\epsilon$ and $D_2^\epsilon$, since the terms containing $D_1^\epsilon$ and $D_2^\epsilon$ vanish in Eq. (\ref{eq:Tcomp}), see Appendix \ref{app:mass_like}. The Einstein summation convention has also been adopted in (\ref{eq:Tcomp}).

In the continuum limit, the four time- and space-dependent angles will code for the components of a curved metric in (1+2)D spacetime. The $e^{\mu}_{(a)}$'s will then code for a $3$-bein or triad, the $e^{(a)}_{\mu}$'s will code will code for its dual 
and $\epsilon$ will go to zero like the temporal and spatial steps of the spacetime lattice. 

\section{Continuum limit} \label{sec:continuous_limit}
To investigate the continuum limit of 
walk (\ref{eq:walk}), we proceed as in \cite{DMD12a, DMD13b, AD15, ADMDB16} and first
interpret $\Psi_{j,p_1,p_2}$ and $\theta^{11}_{j,p_1,p_2}$,  $\theta^{12}_{j,p_1,p_2}$, $\theta^{21}_{j,p_1,p_2}$, $\theta^{22}_{j,p_1,p_2}$ as the values taken by a wave function $\Psi$ and by four functions  $\theta^{11}$,  $\theta^{12}$, $\theta^{21}$, $\theta^{22}$ at spacetime point $(x^0_j = j \epsilon, x^1_{p_1} = p_1 \epsilon/2, x^2_{p_2} = p_2 \epsilon/2)$. The factor $1/2$ is 
necessary to make the continuum limit match with the standard form of the Dirac-equation.
The limit $\epsilon \rightarrow 0$ is then obtained by Taylor expanding (\ref{eq:walk}) at first order in $\epsilon$. The zeroth-order terms cancel each other, and the first-order terms deliver  a 
Schr\"odinger-like equation for $\Psi$ (see Appendix \ref{app:continuous_limit} for a derivation):
\begin{equation} \label{eq:Hamiltonian_Dirac}
i \partial_{0} \Psi = H \Psi \ , 
\end{equation}
where
\begin{equation} \label{eq:Hamiltonian}
H = \sum_{k = 1}^2 \left[ -i \left( B^k \partial_k  + \frac{1}{2}\partial_k B^k \right)\right] + Q \, ,
\end{equation}
with 
\begin{subequations}
\begin{align}
B^k & = e^k_{(a)} {\gamma}^{(0)} {\gamma}^{(a)}  \\
Q & = (m - T_0/4) \, {\gamma}^{(0)} \\
T_0  
& = - \varepsilon^{a b c} \eta_{cd} e_{(a)}^\mu \partial_b e^{(d)}_\mu \, , \label{eq:T_0}
\end{align}
\end{subequations}
and
\begin{align} \label{eq:Clifford_rep}
{\gamma}^{(0)}
& = \left[
    \begin{array}{cc}
       0 & 1 \\
       1 & 0 
    \end{array}
  \right] \nonumber \\
{\gamma}^{(1)} 
&= \left[
    \begin{array}{cc}
       0 & 1 \\
       -1 & 0 
    \end{array}
  \right] \\
{\gamma}^{(2)}
& =  \left[
    \begin{array}{cc}
       i & 0 \\
       0 & -i 
    \end{array}
  \right]. \nonumber
\end{align}

We show in Appendix \ref{app:mass_like_Dirac} that Eq. (\ref{eq:Hamiltonian_Dirac}) is the Hamiltonian form \cite{DOT08} of the Dirac equation for a spin-$1/2$ fermion of mass $m$ in a $(1+2)$D spacetime equipped with metric 
\begin{equation}
g_{\mu\nu}(x^0,x^1,x^2) = \eta_{ab} e^{(a)}_{\mu}(x^0,x^1,x^2) e^{(b)}_{\nu}(x^0,x^1,x^2). 
\end{equation}
The $e^{\mu}_{(a)}$'s are the components of the $3$-bein or triad 
$(e_{(a)}) = (e_{(1)},e_{(2)},e_{(3)})$ on the coordinate basis $(\partial_\mu) = (\partial_0, \partial_1, \partial_2)$ in the tangent space, and 
the $e^{(a)}_\mu$'s are the components of the dual triad
$(e^{(a)})= (e^{(1)},e^{(2)},e^{(3)})$ on the dual coordinate basis $(\partial^\mu) = (\partial^0, \partial^1, \partial^2)$.

All components of the dual triad with one and only one of the indices equal to  the time index $0$ vanish, and $e^{(0)}_0 = 1$, so that:
\begin{equation}
[g_{\mu\nu}] = \left[
\begin{array}{c c c}
1 & 0 & 0 \\
0 & g_{11} & g_{12} \\
0 & g_{21} & g_{22}
\end{array} 
\right],
\label{eq:defg}
\end{equation}
where 
\begin{align}
g_{12}  = g_{21} = &= -e_{2}^{(1)} e_{1}^{(1)} - e_{1}^{(2)} e_{2}^{(2)}  \nonumber \\
g_{11}  &= -(e_{1}^{(1)})^2 - (e_{1}^{(2)})^2  \\
g_{22}  &= -(e_{2}^{(2)})^2 - (e_{2}^{(1)})^2 \,  . \nonumber
\end{align}

Also of interest is the explicit expression of the matrices $B^1$ and $B^2$ in terms of the angles defining the DTQW:
\begin{align}
B^1 &= \left[
\begin{array}{cc}
-\cos \theta^{11} & -i \cos \theta^{12} \\
i \cos \theta^{12} & \cos \theta^{11}
\end{array}
\right]  \nonumber \\
B^2 &= \left[
\begin{array}{cc}
-\cos \theta^{21} & -i \cos \theta^{22} \\
i \cos \theta^{22} & \cos \theta^{21}
\end{array}
\right]  . 
\end{align}
%

\section{DTQWs coupled to GWs}

GWs represent weak gravitational fields propagating in spacetime. Technically, they are particular solutions of Einstein equations linearized around the vacuum. After the proper choice of coordinate system (gauge), these
Einstein equations linearized around the vacuum essentially simplify into Poisson equations for two independent metric components. GWs can thus be expanded in Fourier modes. Hence, of particular interest are the plane GWs. In the so-called traceless gauge
with coordinates $(t, x, y, z)$, the metric of a monochromatic plane GW of pulsation $\omega$ propagating along the $z$ direction takes the standard form $g _{\mu \nu} = \eta_{\mu \nu} + \xi h_{\mu \nu}$ with 
\begin{equation}
h_{\mu \nu} (t, z) = e^{i\omega (t - z)}\left[
\begin{array}{cccc}
0 & 0 & 0 & 0 \\
0 & {\bar F} (t - z) & {\bar G} (t - z) & 0 \\
0 & {\bar G} (t - z)  & -{\bar F} (t - z) & 0 \\
0 & 0 & 0 & 0 
\end{array}
\right].
\end{equation}
Here, the $\eta_{\mu \nu}$'s are the components of the Minkowski metric in inertial coordinates (with signature convention $(+,-,-,-)$), ${\bar F}$ and ${\bar G}$ are two arbitrary 
functions which represent two polarization states, and $\xi$ is a small parameter which traces the perturbative nature of the waves. In other words, $\xi$ is merely introduced as a reminder that the $\xi h_{\mu\nu}$
is considered small with respect to $g_{\mu\nu}$; all results will be henceforth presented as first-order Taylor expansion in $\xi$. The GW affects the metric 
only in planes of constant $z$ parallel to the $(x, y)$ plane and it is therefore customary to study its effect on matter located in these planes. The reduced 
(1+2)D metric in a plane of constant $z$, say $z = 0$, takes the form:
\begin{equation} \label{eq:GWs_metric}
ds^2 = dt^2 - \left(1 - \xi F(t)\right) dx^2 - \left(1 + \xi F(t)\right) dy^2 + 2 \xi G(t)\, dx dy \, ,
\end{equation}
where $F(t) = \exp( i \omega t) {\bar F} (t)$ and $G(t) = \exp( i \omega t) {\bar G} (t)$.

Let $K$ and $K'$ be two arbitrary real constants and perform the following change of coordinates:
\begin{eqnarray}
T & = & t \nonumber \\
X & = & x \, (1- \xi \frac{K}{2}) \\
Y & = & y\, (1- \xi \frac{K'}{2}) \, .\nonumber 
\end{eqnarray}
Since GWs are perturbative solutions of Einstein equations valid only at first order in $\xi$, the components of the metric in the new coordinate system $(T, X, Y)$ need only be computed
at the same order in $\xi$ and one finds:
\begin{align}\label{eq:metricTXY}
ds^2 & =  dT^2 - \left(1 - \xi \left(F(T) - K \right)\right) dX^2 \\
& - \left(1 + \xi \left(F(T) + K'\right)\right) dY^2
 + 2 \xi G(T)\, dX dY \, .  \nonumber
\end{align}

We can choose the triad $(e_{(a)})_{a=1,2,3}$ in such a way that its only non-vanishing components on the coordinate basis $(\partial_T, \partial_X, \partial_Y)$ are $e_{(a = 1)}^X = 1 + \xi \left(F(T)-K\right)/2$, $e_{(a = 2)}^Y = 1 - \xi \left(F(T) + K' \right)/2$, and $e_{(a = 1)}^Y = e_{(a = 2)}^X = \xi G(T)$.
Identify now $(x^0, x^1, x^2)$ introduced in the preceding section with $(T, X, Y)$. A $(1+2)$D DTQW presented in Sec. \ref{sec:first} will simulate, in the continuum limit, a Dirac fermion propagating
in the metric (\ref{eq:metricTXY}), if 
\begin{subequations}
\begin{eqnarray} \label{eq:mapping}
\cos \theta^{11} & = & 1 + \xi \left(F(T) - K \right)/2  \label{eq:theta_11}\\
\cos \theta^{22} & = & 1 - \xi \left(F(T) + K' \right)/2 \label{eq:theta_22}\\
\cos \theta^{12} & = &  \cos \theta^{21} = \xi G(T) \, .
\end{eqnarray}
\end{subequations}
Since $\xi$ is an infinitesimal and we are working at first order in $\xi$, the third equation can be solved immediately by choosing $\theta^{12} = \theta^{21} = \pi/2 - \xi G(T)$. 
Now, if we had not introduced the constants $K$ and $K'$
and had identified directly the coordinates $t$, $x$, and $y$ introduced earlier with the continuum-limit coordinates $x^0$, $x^1$, and $x^2$, Eqs. (\ref{eq:theta_11}) and  (\ref{eq:theta_22}) could not be solved simultaneously, because the absolute value of both cosines must be smaller than one. Let us explain how introducing these two constants enables us to overcome this difficulty. Linear GWs are defined as perturbations of the flat-space time geometry. Treating the metric (\ref{eq:metricTXY}) as a perturbation to the flat Minkowski metric only makes sense if $F(T)$ is bounded.
One can then always find $K$ and $K'$ which make both  $-\left(F(T) - K \right)$ and $\left(F(T) + K' \right)$ positive. Equations (\ref{eq:theta_11}) and  (\ref{eq:theta_22}) are then solved 
by choosing $\theta^{11} = \left(-\xi (F(T) - K) \right)^{1/2}$ and $\theta^{22} = \left( \xi (F(T) + K') \right)^{1/2}$. We will now focus on pure shear GWs, 
for which $F(t) = 0$.
We also retain the simple choice $K=K'=0$.


\section{DTQWs and pure shear GWs} \label{sec:pure_shear_waves}

The DTQWs are entirely defined by the single angle $\theta^{12} = \theta^{21} = \pi/2 - \xi G(T)$, the mass parameter $m$ and the value of $\epsilon$. To make the discussion definite, we set $m = 0$ and $\epsilon =1$. Since $\epsilon=1$, (i) the continuum limit is recovered by considering wave functions which vary only on scales much greater than unity, and (ii) the coordinates used in the continuum limit, $T$, $X$ and $Y$, are related to the lattice integer coordinates, $j$, $p_1=p_X$ and $p_2=p_Y$, by
\begin{equation}
T=T_j=j \, , \ \ X=p_X/2 \, , \ \ Y=p_Y/2 \, .
\end{equation}

Since the advancement operator $V_j$ depends only on the time $T$ and not on $X$ and $Y$, the DTQWs are best analyzed in Fourier space. Let $A(T, X, Y)$ be an arbitrary function defined on the infinite lattice. One can write
(see \cite{AD14}):
\begin{align}
A(T, X, Y) & =  \frac{1}{{\sqrt{2 \pi}}^{\, 2}} \int_{k_X = -\pi}^\pi \int_{k_Y = -\pi}^\pi  {\tilde A}(T, k_X, k_Y) \nonumber \\
&\times   \exp \left(i \left( k_X p_X + k_Y p_Y\right) \right) dk_X dk_Y \, ,
\end{align}
with the Fourier transform on the lattice defined by
\begin{align}
{\tilde A}(T, k_X, k_Y) & =  \frac{1}{\sqrt{2 \pi}^{\, 2}} \sum_{(p_X, p_Y) \in {\mathbb Z}^2} A(T, X, Y) \nonumber \\
&\times   \exp \left(-i \left( k_X p_X + k_Y p_Y\right) \right) \,  .
\end{align}

Each Fourier mode ${\tilde A}(T, k_X, k_Y)$, $(k_X, k_Y) \in [ - \pi, \pi [^2$, is dynamically independent of the other ones.  Its evolution between time $j$ and $j+1$ is fixed by the operator $V_j$ in Fourier representation, that 
depends on $\boldsymbol{k}=(k_X,k_Y)$
through exponentials which describe in Fourier space the translations in physical space.
Since the DTQWs perform pairs of jumps in each direction (see (\ref{eq:1D_block})),
the time-evolution operator in Fourier space is best conceived as a function of $q_X = 2 k_X$ and $q_Y = 2 k_Y$. 
We will denote this operator by $W(\xi, G(T_j); q_X, q_Y)$, making also explicit the dependence with respect to $\xi$ and $G(T_j)$. The wave-vector $\boldsymbol{q}=(q_X,q_Y)$ is actually the one to work with in the continuum-limit because $\boldsymbol{k} \cdot \boldsymbol{p} =\boldsymbol{q}\cdot \boldsymbol{R}$, where $\boldsymbol{p}=(p_X,p_Y)$ and $\boldsymbol{R}=(X,Y)$. 

Note that the Brillouin zone in $k$-space is $[- \pi, + \pi[$ and that the Brillouin zone in $q$-space is therefore $[- 2\pi, + 2\pi[$.

A straightforward computation at first order in $\xi$ delivers $W(\xi, G(T); q_X, q_Y) = W^{(0)}(q_X, q_Y) + \xi G(T) \,  W^{(1)}(q_X, q_Y)$, with
\begin{equation} \label{eq:W0}
W^{(0)}(q_X, q_Y) = \left[
\begin{array}{cc}
e^{i q_X} \cos q_Y & -  e^{-i q_X} \sin q_Y\\
 e^{i q_X} \sin q_Y & e^{-i q_X} \cos q_Y
\end{array}
\right]
\end{equation}
and
\begin{equation}\label{eq:W1}
W^{(1)}(q_X, q_Y) = \left[
\begin{array}{cc}
e^{i q_X} A(q_X, q_Y) & -  e^{-i q_X} B(q_X, q_Y)\\
 e^{i q_X}  B^*(q_X, q_Y)& e^{-i q_X} A^*(q_X, q_Y)
\end{array}
\right],
\end{equation}
where the superscript $*$ denotes complex conjugation, and
\begin{align}
A(q_X, q_Y) &= \frac{1}{\sqrt{2}} \, e^{i\frac{\pi}{4}} \bar{A}(q_X, q_Y) \nonumber \\
B(q_X, q_Y) &= \frac{1}{\sqrt{2}} \, e^{-i\frac{\pi}{4}} \bar{B}(q_X, q_Y) \, ,
\end{align}
with
\begin{align}
\text{Re} \, \bar{A}(q_X,q_Y) &= - \cos(q_X-q_Y) + \cos q_Y - \sin q_Y + \sin 2q_Y \nonumber \\
\text{Re} \, \bar{B}(q_X,q_Y) &=  + \sin(q_X+q_Y) - \sin q_Y + \cos q_Y - \cos 2q_Y \nonumber \\
\text{Im} \, \bar{A}(q_X,q_Y) &= - \cos(q_X+q_Y) + \cos q_Y + \sin q_Y \nonumber \\
\text{Im} \, \bar{B}(q_X,q_Y) &=  + \sin(q_X-q_Y) + \sin q_Y + \cos q_Y - 1 \, .
\end{align}

The operator $W$ controls the entire DTQW dynamics. In particular, at time $T_j$, the energies corresponding to any given $(q_X, q_Y)$ are logarithms of the eigenvalues of $W(\xi, G(T_j); q_X, q_Y)$ and the associated eigenvectors define the eigen-polarizations of the spin-$1/2$ fermion. Thus, a generic
shear GW changes in a time-dependent way the polarization and the energy of the quantum walker. The only modes which are not affected by pure shear GWs are those for which both $A(q_X, q_Y)$ and $B(q_X, q_Y)$ vanish. A direct computation shows that, in the $q$-space Brillouin zone $[- 2 \pi, 2 \pi[$,
both functions vanish simultaneously for two different finite sets of values of $(q_X,q_Y)$. The first set is
$(q_X,q_Y)=2\pi(r_X,r_Y)$ with $(r_X,r_Y)=(0,0)$, $(0,-1)$, $(0,1)$, $(-1,0)$, $(1,0)$, $(-1,-1)$, $(-1,1)$, $(1,-1)$, $(1,1)$, for which the unperturbed (i.e. `free') operator $W^{(0)}$ coincides with unity, so that these modes are unaffected by the free DTQW. The second set is $(q_X,q_Y)=(-\pi/2,\pi/2)+2\pi(s_X,s_Y)$ with $(s_X,s_Y)=(0,0),(0,-1),(1,0),(1,-1)$, for which $W^{(0)}$ coincides with $-i\sigma_1$, so that the spin components of these modes are flipped at each time step by the free DTQW (and shifted by a global phase $-i$). Thus, at first order in the wave perturbation, the modes on which this pure shear GW has no influence correspond to modes for which the probability of presence of the walker, $|\psi^-_{j,p_1,p_2}|^2+|\psi^+_{j,p_1,p_2}|^2$, is unaffected by the free DTQW. All these modes are `small-scale' (i.e. of the same order of magnitude as the lattice spacing) except the $(0, 0)$ mode, which corresponds to a uniform wave function. This mode cannot be populated if space is infinite.

The operator $W^{(1)}(q_X, q_Y)$ can be further characterized by its eigenvalues. They are complex conjugate to each other and have as common modulus $\rho(q_X, q_Y) = \left(| A(q_X, q_Y) |^2 + | B(q_X, q_Y) |^2\right)^{1/2}$. The function $\rho(q_X, q_Y)$ is plotted in Figure 1. It admits four absolute maxima of identical amplitude approximately equal to  $4.69826$. These maxima are located at $(q_X^{a} , q_Y^{b} )$ with $a,b=\pm$ and $q_X^+ \simeq 2.1423$, $q_X^- \simeq -4.14088$, $q_Y^+ \simeq 2.81949$, $q_Y^- \simeq -3.46369$. Generally speaking, the higher $\rho$, the greater the influence of GWs on DTQWs. Thus, pure shear GWs that have a maximal influence on DTQWs on `small' scales comparable to a few lattice steps (a mode with $|\boldsymbol q|=2 \pi$ has a period of $2\pi/|\boldsymbol k|=4\pi/|\boldsymbol q|=2$ lattice steps).

\begin{figure}[h!]
 \includegraphics[width=1\columnwidth]{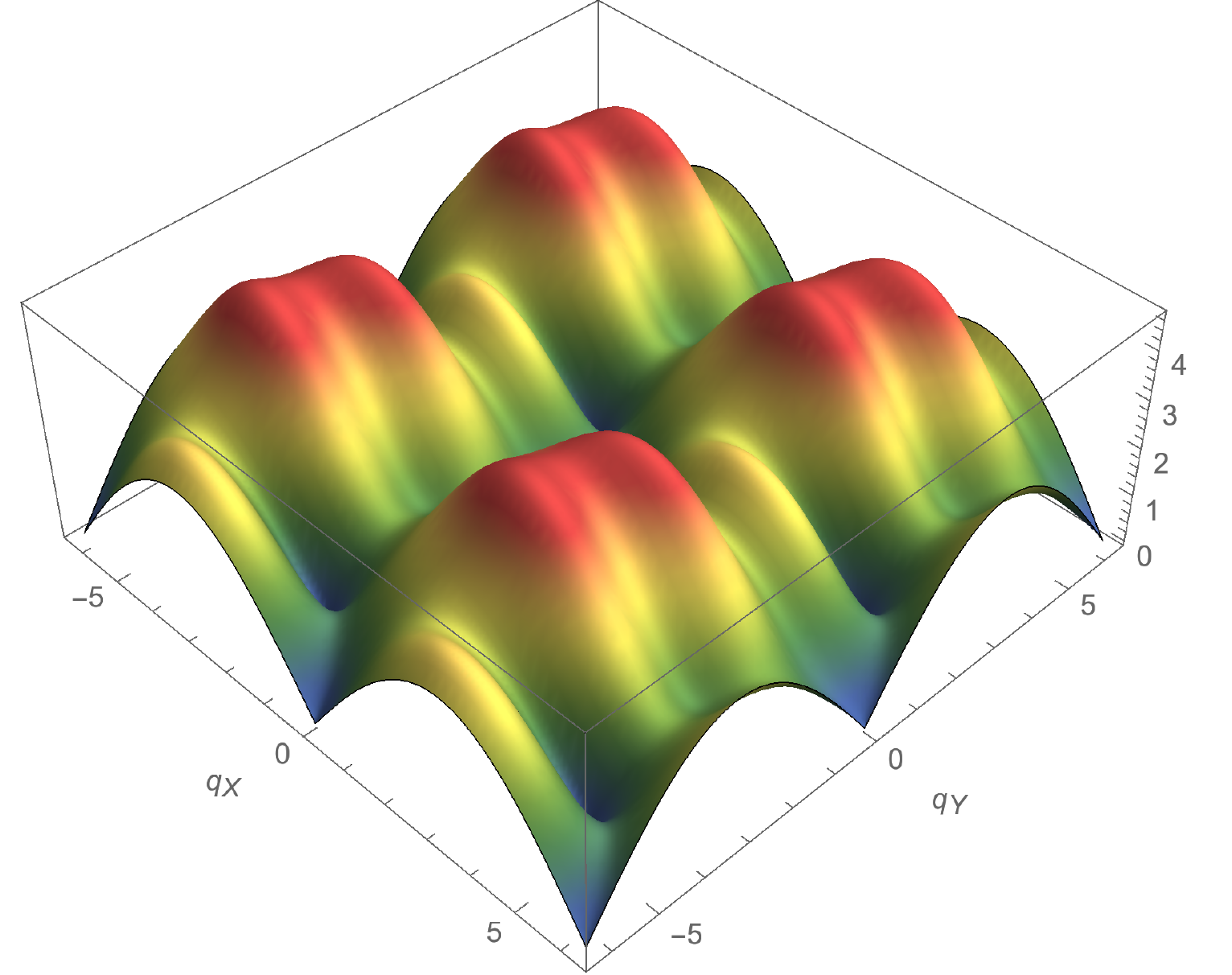}
\caption{(Color online) Modulus  $\rho(q_X, q_Y)$ of the eigenvalues of $W^{(1)}(q_X, q_Y)$.}
\label{fig:Deltaqu}
\end{figure}

We will now present in detail two case-studies. The first one investigates perturbatively how the eigen-energies and eigenvectors behave for large-scale modes, i.e. modes which vary on scales much larger than the lattice 
spacing. The other study is not restricted to large-scale modes, and shows how the interference pattern between two fermion eigen-modes is changed by a GW.

\section{Two case-studies}

\subsection{Large-scale fermion modes} \label{subsec:larg_scale}

Large scales are those relevant to the continuum limit and correspond to values of $q_X$ and $q_Y$ close to zero. One finds, at first order in $q_X$ and $q_Y$:
\begin{align} \label{eq:W_large_scale}
& W(\xi, G(T); q_X, q_Y)  =   \\
&\left[
\begin{array}{cc}
1 + i \left(q_X + \xi G(T) \, q_Y\right)  & - (q_Y + \xi G(T)\, q_X )\\
 q_Y + \xi G(T) \, q_X & 1 - i \left(q_X + \xi G(T) \, q_Y\right) 
\end{array}
\right]. \nonumber
\end{align}
One can check that this is the walk operator obtained in the continuum limit, see Sec. \ref{sec:continuous_limit} and App. \ref{app:continuous_limit}. 

The eigenvalues $\lambda_\pm (\xi, G(T); q_X, q_Y)$ of this operator read, at first order in $\xi$,
\begin{equation}
\lambda_\pm (\xi, G(T); q_X, q_Y) = 1 \mp i\left( 1 + 2 \xi G(T) \frac{q_Xq_Y}{{|\boldsymbol q|}^2} \right) |\boldsymbol q| .
\end{equation}
Associated eigenvectors $V_\pm(\xi, G(T);q_X, q_Y)$ can also be expanded in $\xi$:
\begin{equation}
V_\pm(\xi, G(T);q_X, q_Y) = V^{(0)}_\pm(q_X, q_Y) + \xi G(T) \, V^{(1)}_\pm(q_X, q_Y),
\end{equation}
but exact expressions
are quite involved and depend on the sign of $q_X$. One finds for example
\begin{subequations}
\begin{align}
V^{(0)}_+(q_X, q_Y) & = 
\begin{bmatrix}
\displaystyle{- i \, \frac{q_Y}{q_X + {|\boldsymbol q|}}} \\
1
\end{bmatrix} \\
V^{(1)}_+(q_X, q_Y) & = - i\, \frac{q_X}{q_X + {|\boldsymbol q|}}
\begin{bmatrix}
\displaystyle{q_X - \, \frac{q_Y^2 \left( 1 + 2\, \frac{q_X}{{|\boldsymbol q|}}\right)}{q_X + {|\boldsymbol q|}}}
\\
0
\end{bmatrix} 
\end{align}
\end{subequations}
for positive values of $q_X$.   
 
%

The eigen-energies $E_\pm (\xi, G(T); q_X, q_Y)$ are by definition related to the eigenvalues $\lambda_\pm (\xi, G(T); q_X, q_Y)$ by  $\lambda_\pm (\xi, G(T); q_X, q_Y) = \exp \left( - i E_\pm (\xi, G(T); q_X, q_Y) \right)$. At first order in $|\boldsymbol{q}|$, one finds:
\begin{equation}
E_\pm (\xi, G(T); q_X, q_Y) =  \pm\,  \left( 1 + 2 \xi G(T) \frac{q_X q_Y}{{|\boldsymbol q|}^2} \right) |\boldsymbol q| \, .
\end{equation}

On large scales, the lowest-order effect of the pure shear GWs is thus 
an anisotropic deformation of spatial scales by a factor $\left( 1 - 2 \xi G(T) q_X q_Y /{|\boldsymbol q|}^2 \right)$, the eigen-energies being changed accordingly by the inverse factor, since they are linear in the momentum in free space. 
%

\subsection{Interference pattern modified by GWs} \label{subsec:interference}

We now wish to study the effect of pure shear GWs on all modes, and not only on large-scale ones. GW have been first observed directly using interference with light \cite{Abb16a} but detectors based on interference with matter waves are still considered as a viable alternative (see for example \cite{Gei15a}).
It is thus natural, in the context of the present article, to study the effect of GWs on interference patterns of the walker. 
To be definite, we consider how GWs modify interference patterns between two energy eigen-modes of the free DTQW which share the same energy, thereby ensuring that the interference pattern of these modes is stationary in the absence of GWs.


Consider for example the two wavevectors $(q^1_X = q^1 >0, q^1_Y = 0)$ and $(q^2_X = 0, q^2_Y = q^2 >0)$, and the two initial polarizations
\begin{eqnarray} \label{eq:polars}
\Psi^{1} & = & \left[
\begin{array}{c}
0\\
1
\end{array}
\right] \nonumber \\
\Psi^{2} & = & \left[
\begin{array}{c}
-\frac{i}{\sqrt{2}}\\
\frac{1}{\sqrt{2}}
\end{array}
\right].
\end{eqnarray}
The first is an eigen-polarization of $W^{(0)}(q^1, 0)$, associated to the eigenvalue $\exp(- i q^1)$ and the second is an eigen-polarization of 
$W^{(0)}(0, q^2)$, associated to the eigenvalue $\exp(- i q^2)$.  
In particular, both eigenvalues are the same if $q^1 = q^2 = q$ and the free DTQW then does not modify the interference pattern of the two fermion eigen-modes.
This is not so anymore in the presence of a GW, which changes the interference pattern at each time step $j$ by a contribution proportional to $\xi G(T_j)$. 

Let $N_0(q, X, Y)$ be the initial density of an equal superposition of the two interfering fermion eigen-modes, i.e. $N_0(q, X, Y)=\Psi_0^{\dag}(q,X,Y)\Psi_0(q,X,Y)$, where $\Psi_0(q,X,Y)=\Psi^1\exp[iqX]+\Psi^2\exp[iqY]$, and let 
$N_1(\xi, G(T_{j=1}); q, X, Y)$ be the density after one time step of the DTQW.
The interesting quantity is the relative density variation $\Delta$ after one time-step per unit of $\xi G(T)$,
\begin{align}
&\Delta (q, X, Y) = \\
& \ \ \ \frac{1}{\xi G(T_{j=1})}\, \frac{N_1(\xi, G(T_{j=1}); q, X, Y) - N_0(q, X, Y)}{N_0(q, X, Y)} \, . \nonumber 
\end{align} 
A direct computation shows that 
\begin{equation} \label{eq:initial_density}
N_0 (q, X, Y) = 2 + \sqrt{2} \cos \left( \frac{q}{2} (p_X - p_Y) \right) \, ,
\end{equation}
and
\begin{align}
\Delta (q, X, Y) &= \frac{2 \sqrt 2 }{N_0(q, X, Y)} \cos \left(\frac{q}{2} (p_X - p_Y - 2) \right) \sin^2 q \, .
\end{align}

Figure 2 shows the contours of $N_0(q, X, Y)$ and $\Delta (q, X, Y)$ in the $(X, Y)$ plane for $q = q_{\text{max}}$, for which the effect of the GW is maximum (see the discussion nelow). Both $N_0(q, X, Y)$ and $\Delta (q, X, Y)$ depend on $X$ and $Y$ only through $u = p_X - p_Y = 2(X-Y)$. Figure 3 shows how the profile of $\Delta(q, u)$ as a function of $u$ changes with $q$. The profiles are periodic functions of $u$, and are plotted over two periods.


The net effect of the GW can be measured by $\Delta_M (q) = \mbox{max}_u |\Delta(q, u) |$. This function is $\pi$-periodic in $q$ and also obeys:
\begin{equation} \label{eq:Delta}
\Delta_M(q)= \left\lbrace \begin{array}{ll}
f(q) & \text{if} \ q \in [0,\pi/2[ \\
f(\pi-q) & \text{if} \ q \in [\pi/2,\pi[
\end{array} \right. ,
\end{equation}
with
\begin{align}
f (q) &= \frac{2 \sqrt{2} \sin^2(q) \sqrt{1 - \sin^2(q)/2}}{2 + \sqrt{2} \cos (q) \sqrt{1 - \sin^2(q)/2} - \sin^2(q)} \, .
\end{align}

Figure 4  displays $\Delta_M(q)$ on $[0,\pi[$, which is half of the positive half of the $q$-space Brillouin zone. The absolute minimum is zero and occurs for $q=0$ and $\pi$. The absolute maximum  is approximately $2.48161$ and is reached for 2 values of $q$, $\pi-q_{\text{max}}$ and $q_{\text{max}}\simeq 1.97504$, which correspond respectively to the following two values of the wavelength $\lambda(q) = 2 \pi/ |k| = 4 \pi/ |q|$:  $10.7722$ and  $6.3626$. There is also a local minimum reached for $q=\pi/2$, whose value is $2$.


\begin{figure}
   \begin{minipage}[c]{.4\linewidth}
     \hspace{-1cm}
     \includegraphics[width=1.5 \columnwidth]{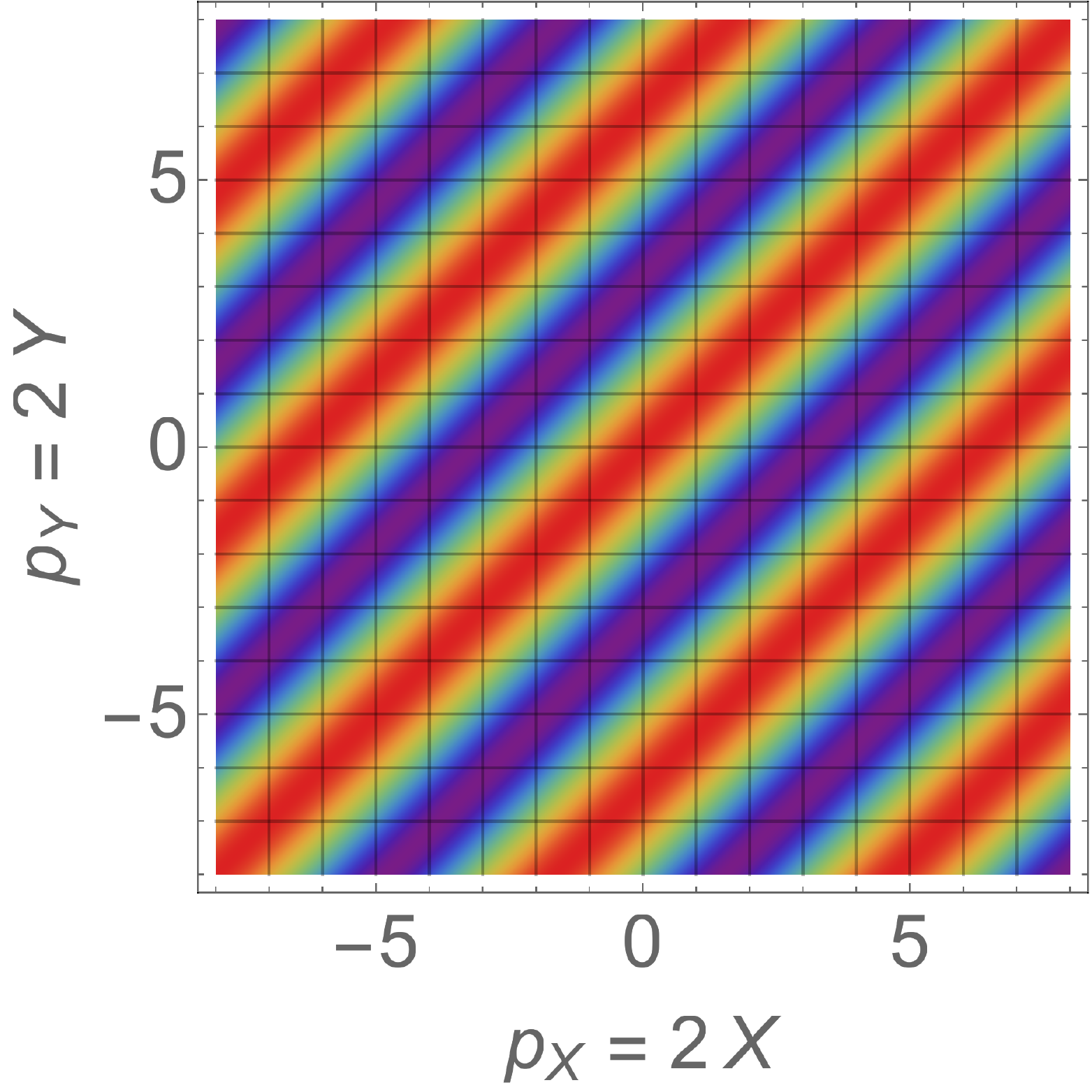}
    \vspace{0.15cm}   
   \end{minipage}
   \hspace{0.4cm}
   \begin{minipage}[c]{.4\linewidth}
     \includegraphics[height = 1.3 \columnwidth, width=0.2\columnwidth]{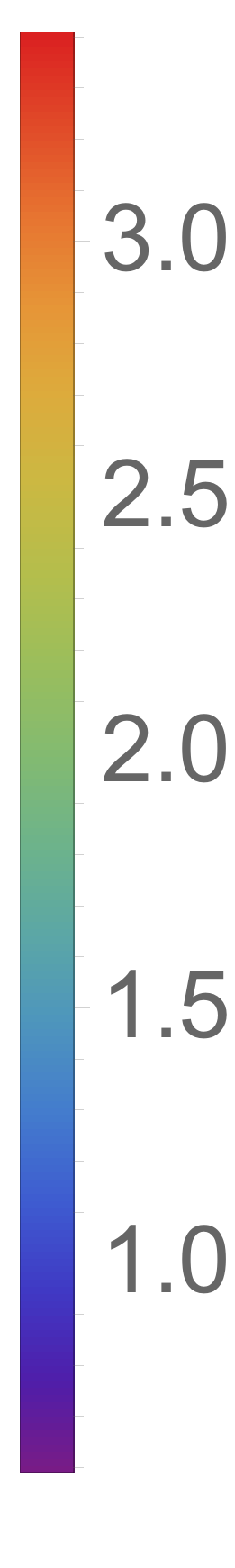}  
   \vspace{0.65cm} \vspace{0.6cm}   \end{minipage} 
     \hspace{6mm}     
   \begin{minipage}[c]{.4\linewidth}
     \includegraphics[width=1.5 \columnwidth]{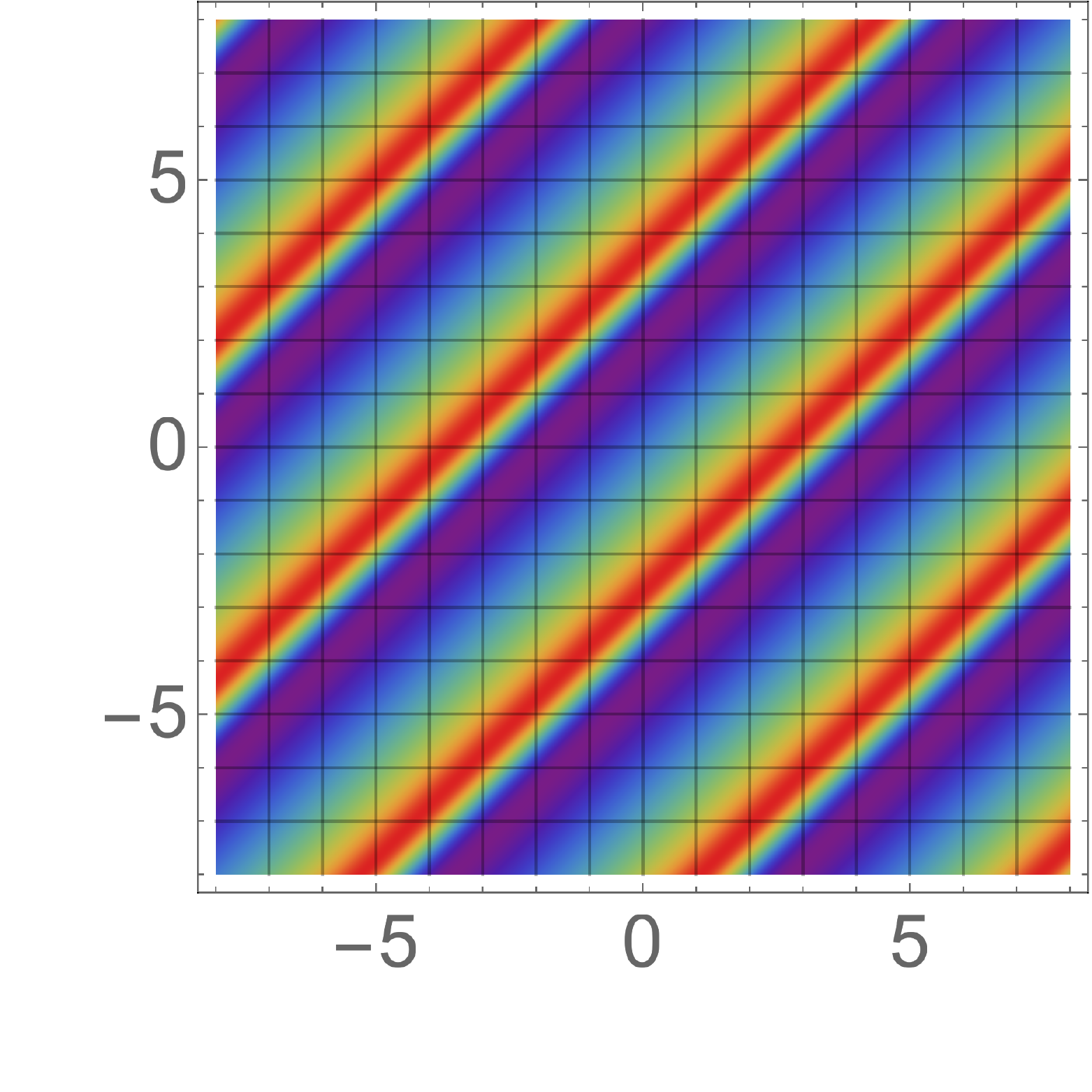} 
   \end{minipage}
   \hspace{0.4cm}
   \begin{minipage}[c]{.4\linewidth}
     \includegraphics[height = 1.3 \columnwidth, width=0.19 \columnwidth]{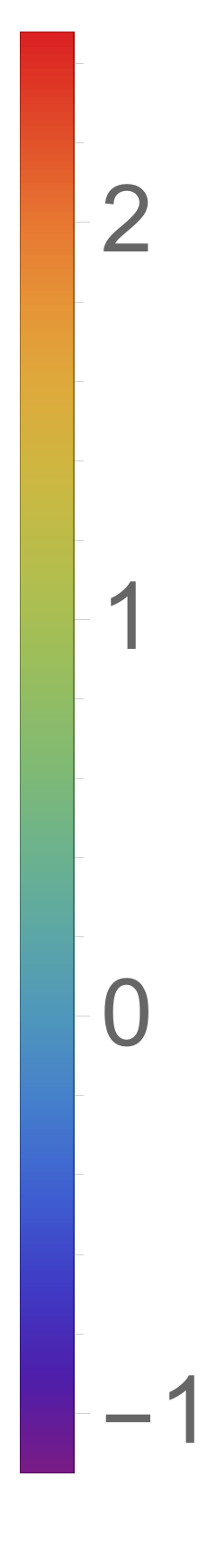}  \vspace{0.8cm} \end{minipage}
\vspace{-0.3cm}     
\caption{(Color online) Density $N_0(q_{\text{max}}, X, Y)$ (top) and relative density variation $\Delta (q_{\text{max}}, X, Y)$ (bottom) in the $(X, Y)$ plane. The mesh represents the 2D lattice on which the DTQW is defined, $p_X$ and $p_Y$ are the integer coordinates used to label the nodes of this lattice, while $X$ and $Y$ are those defined to recover, in the continuum limit, the standard form of the Dirac equation in curved spacetime, Eq. (\ref{eq:Hamiltonian_Dirac}).}
\end{figure}

\begin{figure}[h!]
\includegraphics[width=1\columnwidth]{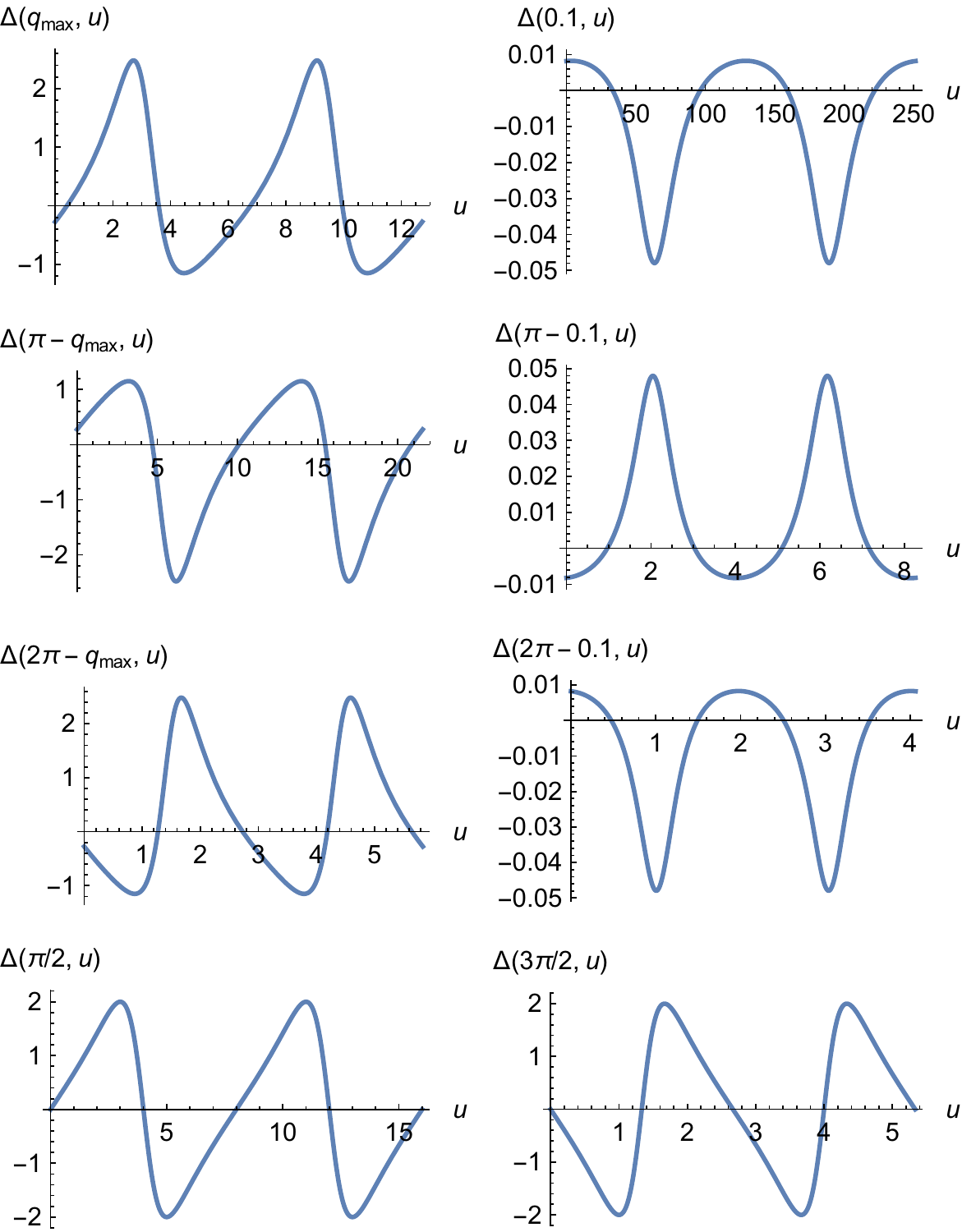}
\caption{(Color online) Profiles of the relative density variation $\Delta(q, u = p_X- p_Y) $ as a function of $u$ for various values of $q$. The profiles are plotted over two rather than one period for a better visual appreciation.}
\label{fig:Deltaqu}
\end{figure}

\begin{figure}[h!]
\includegraphics[width=0.9\columnwidth]{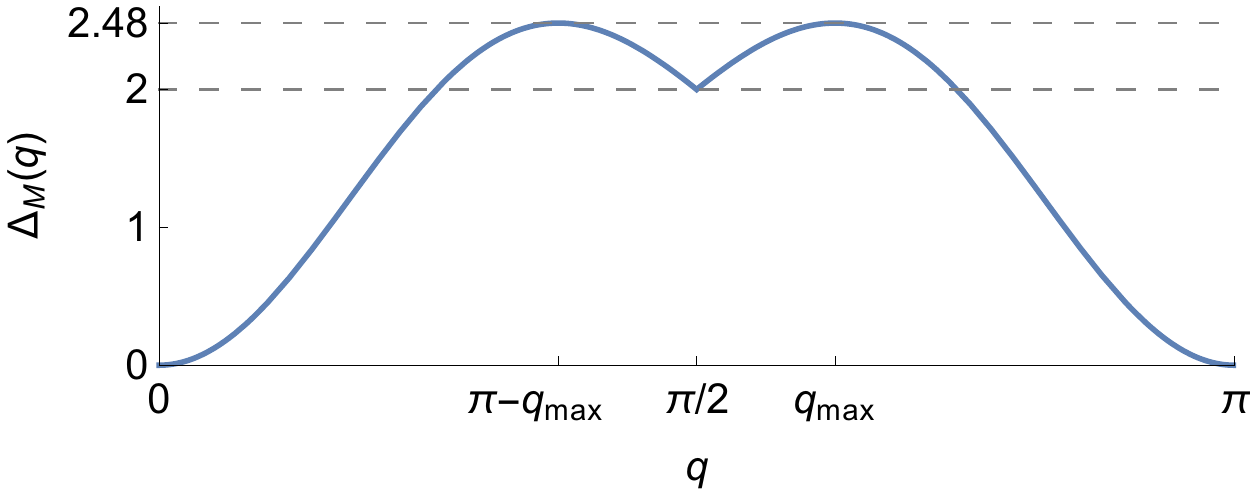} 
\vspace{-0.6cm} \ \ \ \ \ \ \ \ \ \ 
\caption{(Color online) Maximum of the relative density variation induced by a GW on the interference pattern between two fermion eigen-modes, as a function of the common wavevector of these modes, $q$, plotted on half of the positive half of the $q$-space Brillouin zone, namely $[-2 \pi, 2\pi[$.}
\label{fig:DeltaM}
\end{figure}

\section{Discussion}

We have a introduced a new two-dimensional DTQW whose continuum limit coincides with the dynamics of massive spin-1/2 Dirac fermions propagating in a curved spacetime. 
We have then shown how to use these DTQWs to simulate the influence of GWs on Dirac fermions.
We have finally focused on pure shear GWs and investigated in detail how these influence DTQWs on both  large and small scales. On large spatial scales, pure shear GWs locally rescale the eigen-energies anisotropically, to make up for the space deformation induced by the wave, and the eigen-polarizations are modified as well. On smaller scales typically comparable to two or three lattice steps, both polarizations and energies are modified in a non-trivial way; for instance, this has the effect of significatively changing the interference pattern of a superposion of two flat-spacetime eigen-modes.

There exist another way to build DTQWs simulating the interaction of a Dirac field with a gravitational one.
This other construction was presented in \cite{Arrighi_curved_1D_15} for the (1+1)D case 
and an extension to higher spacetime dimensions has just been completed \cite{AF16}. This extension also allows fermions to 
have internal degrees of freedom.

Both constructions use a regular lattice, but they differ in nearly all other aspects. The construction of DTQWs used in this paper is Taylor-made to deliver DTQWs whose continuum limit coincides with the Dirac equation. The other construction is based on Paired Quantum Walks which
admit a whole class of PDEs as continuum limits, including the Dirac equation. As a consequence, each construction has its pros and cons, which we shall now discuss.

As apparent from the above material, our construction works only in synchronous coordinate systems, for which all mixed time-space metric components identically vanish and the time-time component is identical to unity. This does not restrict the gravitational fields one can take into account, since all gravitational fields admit local synchronous coordinates. But it forces the lattice to be regular in precisely these synchronous coordinates, whereas the other construction allows the lattice to be regular in arbitrary coordinates. 

On the other hand, the time-advancement operator $V_j$ used in our construction only involves gauge-invariant  i.e. intrinsic aspects of the gravitational fields, while the time-advancement operator used in the other construction mixes both gauge-invariant and gauge-dependent aspects of the gravitational field. Thus, if one adopts the construction used in this paper, the operator $V_j$ itself can be viewed as a discrete gauge-invariant gravitational field. The gauge dependence of standard continuous physics then manifests itself only in the choice of coordinates when one performs the continuum limit  i.e. in the correspondence between the continuous coordinates $x^{\mu}$ and the lattice indexes. On the other hand, the other construction encodes both the gauge-invariant aspects of the gravitational fields and a choice of coordinates into a single object  i.e. in the time-advancement operator. And one is still free to choose continuous coordinates as one wishes. 

Also, the other construction originally required doubling the number of components of the wave function, but an implementation using only $U(2)$ operators, shifts and swaps has now been presented \cite{AF16}. 

All in all, it seems that both constructions have their own strong and weak points. The construction used in this paper is perhaps more physically grounded and the other one may be of a more mathematical nature.

Let us finally mention of few extensions of this work. The first one is rather straightforward and consists in allowing the operators ${\bf u}$ and ${\bf r}$ to be arbitrary members of $SU(2)$. The extra angles which will appear code for arbitrary discrete electromagnetic fields (see \cite{DMD14}). The other extension 
consists in incorporating in the model extra internal dimensions. This should in particular deliver DTQWs where the time-advancement operator unifies arbitrary discrete Yang-Mills fields with discrete gravitational fields (see \cite{ADMDB16} for (1+1)D DTQWs simulating Dirac fermions in arbitrary Yang-Mills fields). 

In a different direction, the influence of gravitational fields on DTQWs should be systematically studied, with possible applications ranging from fundamental physics to quantum algorithmics.

\vspace{0.5cm}

{\bfseries Aknowledgment.}
Part of this work has been done at the Institute for Molecular Science (second affiliation), and supported by the JSPS summer program (SP16201).

\vspace{0.5cm}

\appendix

\section{Compact form of the mass-like term} \label{app:mass_like}

In this appendix, we show that $T_{\epsilon}(\boldsymbol{\theta})$ given by Eq. (\ref{eq:function_T}) can be put under the form of Eq. (\ref{eq:Tcomp}). 
By definition of the $e_{(a)}^\mu$'s and because $(\eta_{ab}) = \mbox{diag}(0, -1, -1)$,
\begin{eqnarray}
T_\epsilon(\theta) & = & e_{(2)}^k D_0^\epsilon e^{(1)}_k - e_{(1)}^k D_0^\epsilon e^{(2)}_k \nonumber \\
& = & \varepsilon^{0bc} e_{(b)}^k D^\epsilon_0 \eta_{c d} e^{(d)}_k \nonumber\\
& = & \varepsilon^{0bc} e_{(b)}^\mu \eta_{c d} D^\epsilon_0 e^{(d)}_\mu  \nonumber\\
& = & - \varepsilon^{b0c} e_{(b)}^\mu \eta_{c d} D^\epsilon_0 e^{(d)}_\mu.
\end{eqnarray}
where the Einstein summation convention has been used and $\varepsilon^{abc}$ is the completely antisymmetric symbol. The third line is obtained from the second one by taking into account that all $3$-bein and inverse $3$-bein components with one spatial index and the time index $0$ identically vanish. 

Consider now $K^i = \varepsilon^{ibc} e_{(b)}^\mu \eta_{c d} D^\epsilon_i e^{(d)}_\mu$ with $i \in \{1, 2 \}^2$ where the $D^\epsilon_i$ are arbitrary finite 
differences operators which deliver zero when applied to spatially constant objects. 
The quantity $\varepsilon^{ibc}$ is non vanishing only if $b$ or $c$ is equal to zero. The term with $c = 0$ gives a vanishing contribution to $K^i$ because $\eta_{cd}$ then enforces $d = 0$ and the only non-vanishing inverse $3$-bein component $e^{(0)}_\mu$ is
$e^{(0)}_0 = 1$ and $D_0^\epsilon e^{(0)}_0 = 0$.
The only possibly non-vanishing contribution to $K^i$ thus comes form the term where $b = 0$. But the only non-vanishing $3$-bein component $e_{(0)}^\mu$ is $e_{(0)}^0 = 1$. Now, the only non-vanishing inverse $3$-bein component $e^{(d)}_0$ is $e^{(0)}_0 = 1$ and $D_0^\epsilon e^{(0)}_0 = 0$.

The $K^i$'s thus vanish identically and this completes the proof of (\ref{eq:Tcomp}). Note that equation (\ref{eq:Tcomp}) is true independently of the definition of the spatial `discrete derivative' operators $D^\epsilon_1$ and $D^\epsilon_2$.

\section{Mass-like term in the $(1+2)$D Dirac equation} \label{app:mass_like_Dirac}

Consider a spacetime of dimension $(1 + d)$ equipped with metric $g$. 
The Hamiltonian form of the Dirac equation is (see Eq. (35) of \cite{DOT08}):
\begin{equation}
i \partial_0 | \Psi \rangle = \hat{H} | \Psi \rangle \, ,
\end{equation}
with 
\begin{equation}
\hat{H} = \sum_{k=1}^d \left[-i\left( \hat{B}^k \partial_k + \frac{1}{2} \partial_k \hat{B}^k\right) \right] + \hat{M} + \hat{A}(d) \, ,
\end{equation}
where
\begin{align}
\hat{B}^k &= \hat{\alpha}^{(i)} \frac{e^{k}_{(i)}}{e^{0}_{(0)}} + e^{k}_{(0)} \, , \\
\label{eq:mass_term} \hat{M} &= \frac{m}{e^{0}_{(0)}} \hat{\gamma}^{(0)} \, ,
\end{align}
and $\hat{\alpha}^{(k)} = \hat{\gamma}^{(0)}\hat{\gamma}^{(k)}$. As in the main sections of this article, the $e^{\mu}_{(a)}$ are the components of a $(1 + d)$-bein on a coordinate basis.
The components of the inverse $(1 + d)$-bein will be denoted by $e^{(a)}_{\mu}$.
The hat designates operators, as opposed to their representation in a given spin basis. 

In $d=3$, the term $\hat{A}(d)$ reads
\begin{equation} \label{eq:term_A_compact}
\hat{A}(d=3) = - \frac{i}{2} \hat{\gamma}_{5} \hat{\alpha}^{(a)} \tilde{\mathcal{B}}_{(a)} \ ,
\end{equation}
with $\hat{\gamma}_5 = \hat{\gamma}_{(0)} \hat{\gamma}_{(1)} \hat{\gamma}_{(2)} \hat{\gamma}_{(3)}$ and
\begin{equation}
\tilde{\mathcal{B}}_{(a)} = \frac{1}{e^{0}_{(0)}} \left( \frac{1}{2} \varepsilon_{(a) (b) (c) (d)} e^{(b)\mu} e^{(c)\nu} \partial_{\mu} e^{(d)}_{\nu} \right)  , 
\end{equation}
where $\varepsilon_{(a) (b) (c) (d)}$ is the antisymmetric symbol. The notation $\tilde{\mathcal{B}}$ matches the one used in Ref. \cite{DOT08}. In (\ref{eq:term_A_compact}), we can permute $\hat{\gamma}_{5}$ and $\hat{\alpha}^{(a)} = \hat{\gamma}^{(0)} \hat{\gamma}^{(a)}$ because  $\hat{\gamma}_{5}$ anticommutes with all $\hat{\gamma}^{(a)}$'s, and we obtain 
\begin{align} \label{eq:term_A_inverted_gammas}
\hat{A}(d=3) &= - \frac{i}{2} \hat{\gamma}^{(0)} \hat{\gamma}^{(a)} \hat{\gamma}_{5} \\
 \ \  & \times \left[ \frac{1}{e^{0}_{(0)}} \left( \frac{1}{2} \varepsilon_{(a) (b) (c) (d)} e^{(b)\mu} e^{(c)\nu} \partial_{\mu} e^{(d)}_{\nu} \right) \right]  ,  \nonumber
\end{align}
where $\times$ denotes the multiplication. After a cyclic, hence odd permutation of the four indices, (\ref{eq:term_A_inverted_gammas}) reads
\begin{align} \label{eq:term_A_final_dim3}
\hat{A}(d=3) &= -\frac{\hat{\gamma}^{(0)}}{8e^{0}_{(0)}} \\
& \times \underbrace{\left[ -2i \varepsilon_{ (b) (c) (d) (a)} \hat{\gamma}^{(a)} \hat{\gamma}_{5} \right]}_{\hat{J}_{(b)(c)(c)}(d=3)}  e^{(b)\mu} e^{(c)\nu} \partial_{\mu} e^{(d)}_{\nu}    . \nonumber
\end{align} 
The expression of $\hat{J}_{(b)(c)(d)}$ given above is only valid in $(1+ 3)$ dimensional spacetimes. In arbitrary dimension $(1 + d)$, it reads (Eq. ($25$-$c$) of \cite{DOT08}):
\begin{equation}
\hat{J}_{(b)(c)(d)} = \{ \hat{\gamma}_{(b)}, \hat{S}_{(c)(d)} \} \, ,
\end{equation}
where
\begin{equation}
\hat{S}_{(c)(d)} = \frac{i}{2} [\hat{\gamma}_{(c)}, \hat{\gamma}_{(d)}] \, .
\end{equation}
Now, $\hat{J}_{(b)(c)(d)}$ is obviously antisymmetric in $(c,d)$, but we can show by an explicit computation that it is also antisymmetric in $(b,c)$, which is a consequence of the Clifford algebra satisfied by the gamma matrices. Hence, if one of the indices is repeated, $\hat{J}_{(b)(c)(d)}$ vanishes. This shows that $\hat{J}_{(b)(c)(d)}$ vanishes in (1+1)D spacetimes, so that
\begin{equation}
\hat{A}(d=1) = 0   \, . 
\end{equation}
In (1+2) dimension, there are $6$ non vanishing components of $\hat{J}_{(b)(c)(d)}$:
\begin{align}
\hat{J}_{012} & = - \hat{J}_{102} \nonumber  \\
\hat{J}_{201} & = - \hat{J}_{021} \\
\hat{J}_{120} & = - \hat{J}_{210} \, . \nonumber 
\end{align}
In representation (\ref{eq:Clifford_rep}), we can check that
\begin{equation} \label{eq:J_tensor_components}
J_{012} = J_{201} = J_{120} = 2 \times \mathbf{1}_2 = 2 \, ,
\end{equation}
where $\mathbf{1}_2$ is the $2 \times 2$ identity matrix. The previous equalities are then valid in any representation (trivial to check). Hence, a compact and generic expression for $\hat{J}_{(\beta)(\rho)(\sigma)}$ in $d=2$ is:
\begin{equation} \label{eq:J_in_dim2}
\hat{J}_{(b)(c)(d)} = 2 \varepsilon_{(b)(c)(d)} \, ,
\end{equation}
where $\varepsilon_{(b)(c)(d)}$ is the $3$D antisymmetric symbol. Thus, a generic expression for the term $\hat{A}(d)$ in (1+2) dimensions is
\begin{equation}
\hat{A}(d=2) = -\frac{1}{4e^{0}_{(0)}} \underbrace{ \left(  \varepsilon_{ (b) (c) (d)} \underbrace{e^{(b)\mu} e^{(c)\nu} \partial_{\mu} e^{(d)}_{\nu} }_{\tau_{(b)(c)(d)}} \right)}_{T_0} \hat{\gamma}^{(0)} \,  .
\end{equation}
Observe that $\hat{A}(d=2)$ has, in the spin Hilbert space, the same structure as the mass term (\ref{eq:mass_term}). We have
\begin{equation}
T_0 = \tau^{012} - \tau^{102} + \tau^{201} - \tau^{021} + \tau^{120} - \tau^{210} \, .
\end{equation}
If we take a dual $3$-bein of the form
\begin{equation} \label{eq:spatial_sheet_3bein}
\left[ e^{(\alpha)}_{\mu} \right] = \left[
\begin{array}{ccc}
1 & 0 & 0 \\
0 & e^{(1)}_1 & e^{(1)}_2 \\
0 & e^{(2)}_1 & e^{(2)}_2
\end{array}
\right]  ,
\end{equation}
then
\begin{equation} \label{eq:T_space_time_decompsition}
T_0 = e^{(1)\nu} \partial_0 e^{(2)}_{\nu} - e^{(2)\nu} \partial_0 e^{(1)}_{\nu} \, ,
\end{equation}
which, after a small calculation (the same as that of App. \ref{app:mass_like} but replacing the $D_b$'s by $\partial_b$'s), can be shown to coincide with Eq. (\ref{eq:T_0}). This finalizes the proof that the (1+2)D Dirac equation in representation (\ref{eq:Clifford_rep}) can be put under the form of Eq. (\ref{eq:Hamiltonian_Dirac}).


\section{Detailed computation of the continuum limit} \label{app:continuous_limit}

We are going to show that the first-order expansion of the walk operator reads $V_j=1-i\epsilon H$ with $H$ given by (\ref{eq:Hamiltonian}). 

We define $\mathcal{M}_{\epsilon}=m-T_{\epsilon}(\boldsymbol{\theta})/4$. At first order in $\epsilon$,
\begin{equation} \label{eq:first_order_mass}
\mathbf{Q} \left( \epsilon\mathcal{M}_{\epsilon}\right) = 1 - i\epsilon\mathcal{M}_0 \sigma_1 \, ,
\end{equation}
where $\mathcal{M}_0$ is the limit of $\mathcal{M}_{\epsilon}$ when $\epsilon$ goes to zero, and $\sigma_1$ the first Pauli matrix.

The spin-dependent shift operators can be written
\begin{equation}
\mathbf{S}_k = e^{i\epsilon P_k \sigma_3}
\, , 
\end{equation}
with the momentum operator
\begin{equation}
P_k = -i\partial_k \, .
\end{equation}
Thus
\begin{equation}
\mathbf{S}_k = 1 + i\epsilon P_k \sigma_3 \, , 
\end{equation}
from which one obtains
\begin{equation} \label{eq:first_order_1Dwalk}
\mathbf{W}_{k}(\theta_j) = 1 - i \epsilon \mathbf{H}_k(\theta_j) \, ,
\end{equation}
where the `1D Hamiltonians' $\mathbf{H}_k(\theta_j)$ read
\begin{equation}
\label{eq:C8}
\mathbf{H}_k(\theta_j) = -\frac{1}{2}\mathbf{R}^{-1}(\theta_j) \mathbf{U}(\theta_j) \, \lbrace \mathbf{U}(\theta_j),P_k(\sigma_3 \rbrace \, \mathbf{R}(\theta_j) \star \, .
\end{equation}
Here $\lbrace A,B \rbrace = AB+BC$ and the notation $P_k(O_j\star$ means that $P_k$ applies to $O_j\Psi_j$ and not only to $O_j:(j,p_1,p_2)\mapsto O_{j,p_1,p_2}$, which is any operator acting on $\Psi_j$, i.e.
\begin{equation}
P_k(O_j\star \Psi_j \equiv P_k(O_j  \Psi_j) \, ,
\end{equation}
so that $\mathbf{H}_k(\theta_j) \Psi_j$ reads
\begin{align}
\mathbf{H}_k(\theta_j) \Psi_j =& -\frac{1}{2}\mathbf{R}^{-1}(\theta_j) \mathbf{U}(\theta_j)  \mathbf{U}(\theta_j) \, P_k \! \left(  \sigma_3 \mathbf{R}(\theta_j) \Psi_j \right) \nonumber \\
  &-\frac{1}{2}\mathbf{R}^{-1}(\theta_j) \mathbf{U}(\theta_j) \,  P_k  \! \left( \sigma_3 \mathbf{U}(\theta_j) \mathbf{R}(\theta_j) \Psi_j \right) \, .
\end{align}

The previous expressions lead to
\begin{equation}
V_j = 1 - i\epsilon \mathbf{H} \, ,
\end{equation}
with
\begin{equation}
\mathbf{H} = \sum_{k=1}^2 \mathbf{K}(\theta_j^{k1},\theta_j^{k2})  +  \mathcal{M}_0 \sigma_x \, ,
\end{equation}
where
\begin{equation}
\mathbf{K}(\theta_j^{k1},\theta_j^{k2})= \mathbf{H}_k(\theta_j^{k1}) + \mathbf{\Pi}^{-1} \mathbf{H}_k(\theta_j^{k2}) \mathbf{\Pi} \, .
\end{equation}

We will now show that $\mathbf{H}$ is identical to $H$ given by (\ref{eq:Hamiltonian}).
We can immediately check that the mass term $\mathcal{M}_0 \sigma_x$ is the one given by (\ref{eq:mass_term}). We therefore need to show that $\mathbf{K}(\theta_j^{k1},\theta_j^{k2})$ is identical to the term in square brackets from Eq. (\ref{eq:Hamiltonian}).
The total Hamiltonian $\mathbf{H}$ is linear in the $\mathbf{H}_k(\theta_j^{kl})$. It is convenient to compute first each $\mathbf{H}_k(\theta_j^{kl})$ separately and then combine all results into $\mathbf{H}$.
For each given $(j, k, l)$, one can safely use the simplified notations
\begin{align}
U = \mathbf{U}(\theta_j^{kl}) \ \ \ \ \mathrm{and} \ \ \ \ \ 
R = \mathbf{R}(\theta_j^{kl}) \, .
\end{align}
Equation (\ref{eq:C8}) leads to
\begin{align}
\mathbf{H}_k(\theta_j^{kl}) &= -\frac{1}{2}R^{-1} U \left[ U P_k(\sigma_3  + P_k(\sigma_3 U  \right] R \star\\
&= \frac{i}{2} R^{-1} U \left[ U \partial_k(\sigma_3 R \star + \partial_k(\sigma_3 UR \star \right] \nonumber \, ,
\end{align}
which delivers
\begin{equation}
\mathbf{H}_k(\theta_j^{kl}) = \frac{i}{2}R^{-1}U \left[ \lbrace U,\sigma_3 \rbrace \left( R\partial_k + \partial_k R\right) + \sigma_3 (\partial_k U) R \right] \, .
\end{equation}
Now, we have
\begin{equation}
\lbrace U,\sigma_3 \rbrace = - 2 c \mathbf{1}_2 \, ,
\end{equation}
where $\mathbf{1}_2$ is the $2\times 2$ identity matrix and
\begin{equation}
c = \cos \theta_j^{kl} \, ,
\end{equation}
and 
\begin{equation}
R^{-1}UR =- \sigma_3 \, .
\end{equation}
Hence,
\begin{equation}
\mathbf{H}_k(\theta_j^{kl}) = -i \left( -c \sigma_3 \partial_k + \frac{1}{2} \Omega  \right) \, ,
\end{equation}
with
\begin{equation}
\Omega = R^{-1}U \left( 2 c \partial_k R - \sigma_3 (\partial_k U) R \right) \, .
\end{equation}
We have
\begin{equation}
-c\sigma_3 = D^{kl} = \begin{bmatrix}
-\cos \theta_j^{kl} & 0 \\ 0 & \cos \theta_j^{kl}
\end{bmatrix} \, ,
\end{equation}
and a straightforward computation involving only matrix products and derivations shows that
\begin{equation}
\label{eq:nosum1}
\Omega = \partial_kD^{kl} \, ,
\end{equation}
so that
\begin{equation}
\label{eq:nosum2}
\mathbf{H}_k(\theta_j^{kl}) = -i \left( D^{kl} \partial_k + \frac{1}{2} \partial_k D^{kl} \right) \, .
\end{equation}
Now, we have
\begin{equation}
\label{eq:nosum3}
\mathbf{\Pi}^{-1}\mathbf{H}_k(\theta_j^{kl}) \mathbf{\Pi}= -i \left( A^{kl} \partial_k + \frac{1}{2} \partial_k A^{kl} \right) \, ,
\end{equation}
where the antidiagonal matrix is given by
\begin{equation}
\label{eq:nosum4}
A^{kl} = \mathbf{\Pi}^{-1} D^{kl} \mathbf{\Pi} = \begin{bmatrix}
0 & -i \cos \theta_j^{kl} \\ i \cos \theta_j^{kl} & 0 
\end{bmatrix} \, .
\end{equation}
Note that equations (\ref{eq:nosum1})-(\ref{eq:nosum4}) are all written at fixed $(j, k, l)$ and thus involve in particular no summation over $k$.

Combining (\ref{eq:nosum1}) and (\ref{eq:nosum4}) leads to
\begin{equation}
\sum_{k=1}^2 \mathbf{K}(\theta_j^{k1},\theta_j^{k2}) = -i\left( B^k \partial_k + \frac{1}{2} \partial_k B^k \right) \, , 
\end{equation}
where summation over \emph{is} implied in the right-hand side. This completes the proof.

\end{document}